\documentclass[apjl]{emulateapj}
\usepackage{psfig,amsfonts,amsmath,graphicx,natbib,apjfonts,lscape}

\def\sw{Sw\,1644+57}
\def\chandra{{\it Chandra}}
\def\swift{{\it Swift}}

\def\cfa{1}
\def\ami{2}
\def\hu{3}
\def\mpifr{4}

\def\york{6}
\def\hrao{7}

\begin{document}

\title{Radio Monitoring of the Tidal Disruption Event
Swift\,J164449.3+573451. II.  The Relativistic Jet Shuts Off and a
Transition to Forward Shock X-ray/Radio Emission}

\author{ 
B.~A.~Zauderer\altaffilmark{\cfa},
E.~Berger\altaffilmark{\cfa},
R.~Margutti\altaffilmark{\cfa},
G.~G.~Pooley\altaffilmark{\ami},
R.~Sari\altaffilmark{\hu},
A.~M.~Soderberg\altaffilmark{\cfa},
A.~Brunthaler\altaffilmark{\mpifr},
and M.~F.~Bietenholz\altaffilmark{\york,}\altaffilmark{\hrao}
}

\altaffiltext{\cfa}{Harvard-Smithsonian Center for Astrophysics, 60
Garden Street, Cambridge, MA 02138}

\altaffiltext{\ami}{Mullard Radio Observatory, Cavendish Laboratory,
Cambridge, CB3 0HE UK}

\altaffiltext{\hu}{Racah Institute of Physics, The Hebrew University,
91904 Jerusalem, Israel}

\altaffiltext{\mpifr}{Max-Planck-Institut f$\ddot{\rm u}$r
Radioastronomie, Auf dem H$\ddot{\rm u}$gel 69, 53121 Bonn, Germany}

\altaffiltext{\york}{Department of Physics and Astronomy, York
University, Toronto, Ontario, Canada}

\altaffiltext{\hrao}{Hartebeesthoek Radio Astronomy Observatory, PO
Box 443, Krugersdorp, 1740 South Africa}

\begin{abstract} We present continued multi-frequency radio
observations of the relativistic tidal disruption event
Swift\,J164449.3+573451 (Sw\,1644+57) extending to $\delta t\approx
600$ d.  The data were obtained with the JVLA and AMI Large Array as
part of our on-going study of the jet energetics and the density
structure of the parsec-scale environment around the disrupting
supermassive black hole (SMBH).  We combine these data with public
{\it Swift}/XRT and {\it Chandra} X-ray observations over the same
time-frame to show that the jet has undergone a dramatic transition
starting at $\approx 500$ d, with a sharp decline in the X-ray flux by
about a factor of 170 on a timescale of $\delta t/t\lesssim 0.2$ (and
by a factor of 15 in $\delta t/t\approx 0.05$).  The rapid decline
rules out a forward shock origin (direct or reprocessing) for the
X-ray emission at $\lesssim 500$ d, and instead points to internal
dissipation in the inner jet.  On the other hand, our radio data
uniquely demonstrate that the low X-ray flux measured by {\it Chandra}
at $\approx 610$ d is consistent with emission from the forward shock.
Furthermore, the {\it Chandra} data are inconsistent with thermal
emission from the accretion disk itself since the expected temperature
of $\sim 30-60$ eV and inner radius of $\sim 2-10\,R_s$ cannot
accommodate the observed flux level or the detected emission at
$\gtrsim 1$ keV.  We associate the rapid decline with a turn off of
the relativistic jet when the mass accretion rate dropped below $\sim
\dot{M}_{\rm Edd} \approx 0.006$ M$_\odot$ yr$^{-1}$ (for a $3\times
10^6$ M$_\odot$ black hole and order unity efficiency) indicating that
the peak accretion rate was about $330\,\dot{M}_{\rm Edd}$, and the
total accreted mass by $\delta t\approx 500$ d is about $0.15$
M$_\odot$.  From the radio data we further find significant flattening
in the integrated energy of the forward shock at $\delta t\gtrsim 250$
d with $E_{\rm j,iso}\approx 2\times 10^{54}$ erg ($E_j\approx
10^{52}$ erg for a jet opening angle, $\theta_j=0.1$) following a rise
by about a factor of 15 at $\approx 30-250$ d.  Projecting forward, we
predict that the emission in the radio and X-ray bands will evolve in
tandem with similar decline rates.  \end{abstract}


\section{Introduction}
\label{sec:into}

The unusual $\gamma$-ray/X-ray transient \sw\ has been broadly
interpreted as the first example of a tidal disruption event (TDE)
powering a relativistic jet (e.g.,
\citealt{bgm+11,bkg+11,ltc+11,zbs+11}).  As such, \sw\ provides unique
insight into the formation (and potentially termination) of
relativistic jets in supermassive black holes, a process that is not
observed in active galactic nuclei due to long lifetimes of $\gtrsim
10^7$ yr.  One of the primary observations supporting the TDE
relativistic jet scenario in \sw\ is the long-term evolution of the
X-ray light curve, roughly following a $t^{-5/3}$ power law decline
\citep{bkg+11} as expected for the fallback rate of tidally disrupted
material (e.g., \citealt{ree88}).  In addition, the mean X-ray
luminosity at early time, $L_{\rm X,iso}\approx 10^{47}$ erg s$^{-1}$
(flaring to $\approx 3\times 10^{48}$ erg s$^{-1}$ on a $\sim 10^2$ s
timescale; \citealt{bkg+11}), exceeded the Eddington limit of a $\sim
10^6-10^7$ M$_\odot$ black hole by about $2-3$ orders of magnitude,
supporting the presence of a collimated relativistic outflow.
Independently, our discovery of bright radio synchrotron emission from
\sw\ established the presence of a relativistic outflow with a Lorentz
factor of $\Gamma\sim {\rm few}$, launched at the same time as the
onset of $\gamma$-ray emission \citep{zbs+11,bzp+12}.  The basic
picture, therefore, is of X-ray emission likely from internal
dissipation in the inner part of the jet (at $r\sim 10^{15}-10^{16}$
cm) and radio emission from the expanding forward shock (at $r\sim
10^{18}-10^{19}$ cm).

While the formation of relativistic jets was not predicted in TDE
models, a super-Eddington accretion phase was expected (e.g.,
\citealt{ek89,ulm99,sq09}) and the potential for jets was discussed
\citep{gm11}.  The latter paper considered two distinct possibilities
for jet formation, during the super-Eddington phase, or at a later
time when the accretion rate drops below a few percent of the
Eddington rate (motivated by observations of steady jets in X-ray
binaries).  The rapid formation of the relativistic jet in \sw\ points
to the former scenario.  The peak mass accretion rate and duration of
the super-Eddington phase are expected to depend on the mass of the
black hole, with $\dot{M}_p\approx 1.4$ M$_\odot$ yr$^{-1}$ and
$\tau_{\rm SE}\approx 1.5$ yr for a $3\times 10^6$ M$_\odot$ black
hole (e.g., \citealt{ek89,cgn+12}), the inferred mass of the
disrupting supermassive black hole in \sw\
\citep{bgm+11,bkg+11,ltc+11,zbs+11}.  Although it is unclear what, if
anything, happens to a TDE jet when the accretion declines below the
Eddington limit, an analogy with X-ray binaries indicates that
relativistic jet ejections will likely be restricted to the
super-Eddington phase (e.g., \citealt{fct+99,cgn+12}).

To take advantage of this unique opportunity to study the birth and
evolution of a relativistic jet from a supermassive black hole, and to
track the jet properties of a TDE, we have been carrying out a
long-term monitoring campaign of the radio emission from \sw, in
conjunction with X-ray data \citep{zbs+11,bzp+12}.  Here we present
new radio observations that extend to $\delta t\approx 600$ d, and use
these data to determine the continued evolution of the integrated
forward shock energy.  We combine these measurements with public
\swift/XRT and \chandra\ observations over the same timescale to show
that the relativistic jet has shut off at $\delta t\approx 500$ d,
marked by a steep decline in the X-ray luminosity
\citep{atel4398,atel4610}.  The radio data allow us to uniquely
determine that the X-ray flux measured in the \chandra\ data is
consistent with emission from the forward shock; a model of thermal
emission from the accretion disk can be ruled out by the flux and
spectrum of the X-ray emission.  Associating the rapid decline with
the timescale at which $\dot{M}\approx \dot{M}_{\rm Edd}$, we infer
the peak mass accretion rate and the total accreted mass at $\delta
t\lesssim 500$ d.

\section{Radio Observations}
\label{sec:radio_obs}

Previous radio observations of \sw\ extending to $\delta t\approx 26$
d were presented in \citet{zbs+11}, while data extending to $\delta
t\approx 216$ d were presented in \citet{bzp+12}; hereafter, Paper I.
Here we report new observations extending to $\delta t\approx 600$ d.
All times are measured relative to a $\gamma$-ray onset date of 2011
March 25.5 UT.  Throughout the paper we use the standard cosmological
constants with $H_0=70$ km s$^{-1}$ Mpc$^{-1}$, $\Omega_m=0.27$ and
$\Omega_\Lambda=0.73$.

We observed \sw\ with the Karl G.~Jansky Very Large Array
(JVLA\footnotemark\footnotetext{The JVLA is operated by the National
Radio Astronomy Observatory, a facility of the National Science
Foundation operated under cooperative agreement by Associated
Universities, Inc.  The observations presented here were obtained as
part of programs 11A-266 and 12A-280.}) using the Wideband
Interferometric Digital Architecture (WIDAR; \citealt{pcb+11})
correlator to obtain up to 2 GHz of bandwidth at several frequencies.
At all frequencies we used 3C286 for bandpass and flux calibration,
while phase calibration was performed using J1634+6245 at 1.8 GHz and
J1638+5720 at all other frequencies.  We reduced and imaged the data
with the Astronomical Image Processing System (AIPS; \citealt{gre03})
software package.  The observations are summarized in
Table~\ref{tab:data}.

We also observed \sw\ with the AMI Large Array (AMI-LA) at 15.4 GHz
with a bandwidth of 3.75 GHz using J1638+5720 for phase calibration
and 3C48 and 3C286 for flux calibration.  The AMI-LA observations are
summarized in Table~\ref{tab:data}.

\section{X-ray Observations}
\label{sec:xray_obs}

\begin{figure}
\epsscale{1}
\plotone{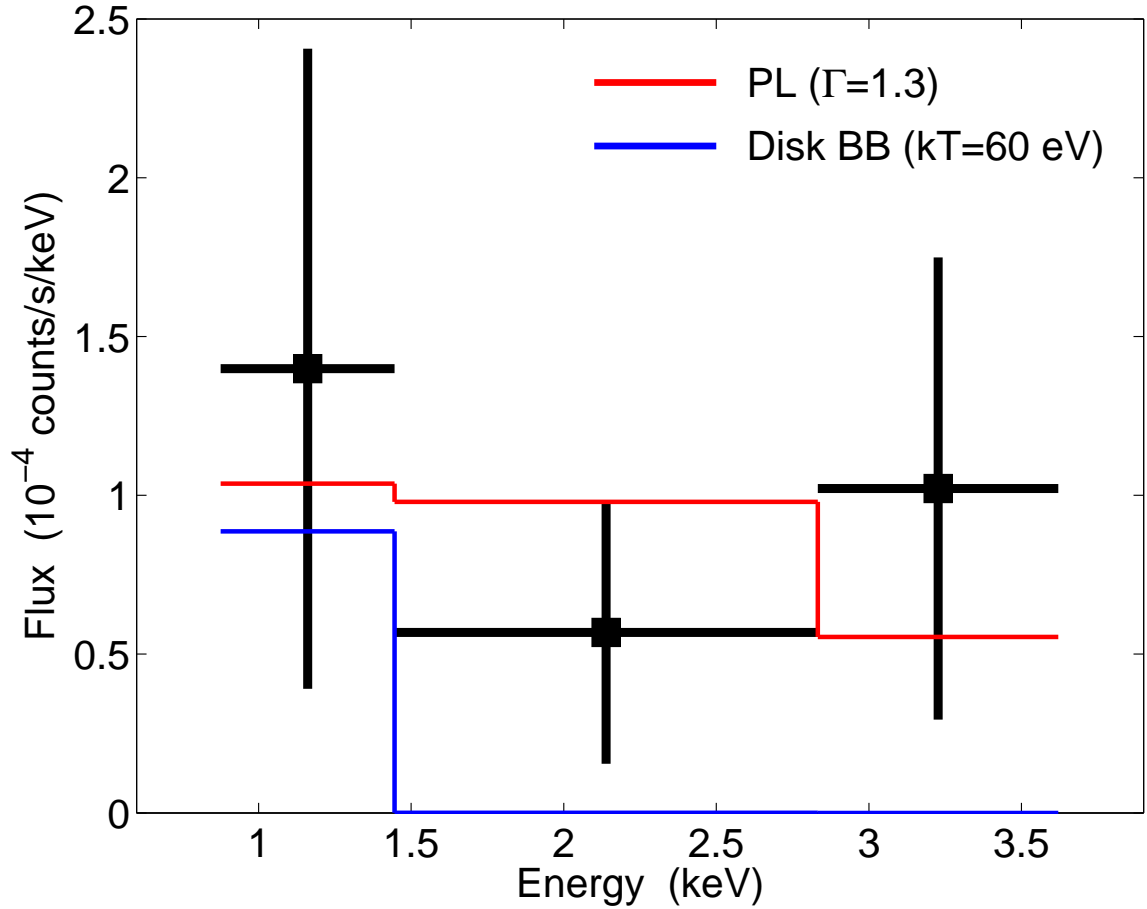}
\caption{Spectrum of the X-ray emission from the \chandra\ observation
at $\delta t\approx 610$ d (black points).  Also shown are the best
fit power law model (red line), and a multi-temperature disk blackbody
model with $kT\approx 60$ eV, appropriate for an accretion disk with
an inner radius of $2\,R_s$ around a $3\times 10^6$ M$_\odot$ black
hole (blue line).  The disk model provides a poor fit to the data at
$\gtrsim 1$ keV.  In addition, to fit the flux at $\sim 1$ keV this
model requires a radius of $3.4\times 10^{13}\,{\rm cm}\approx
40\,R_s$, which is inconsistent with the temperature.  We therefore
conclude that the X-ray emission at late time is not due to the
accretion disk.
\label{fig:xspec}} 
\end{figure}

\chandra/ACIS-S observations of \sw\ (PI: Tanvir; \citealt{atel4610})
started on 2012 November 26.42 UT ($\delta t\approx 610$ d), with a
total exposure time of 24.7 ks.  We analyzed the public data with the
CIAO software package (v4.4), using the calibration database CALDB
(v4.5.3) and standard ACIS data filtering.  Using {\tt wavedetect} we
detect \sw\ at a significance level of $2.8\sigma$ with a count rate
of $(2.0\pm 0.9)\times 10^{-4}$ count s$^{-1}$ ($0.5-8$ keV; $1.5''$
radius aperture).  We note that emission is detected with a roughly
flat distribution in counts s$^{-1}$ keV$^{-1}$ at $\approx 1-3.5$ keV
(Figure~\ref{fig:xspec}).

To convert the observed count rate to a flux we note that starting at
$\delta t\approx 23$ d the X-ray emission from \sw\ undergoes spectral
hardening\footnotemark\footnotetext{{\tt
http://www.swift.ac.uk/burst\_analyser/00450158}}, with the photon
index evolving to $\Gamma\approx 1.3$ at $\delta t\gtrsim 230$ d.  We
therefore use an absorbed power law spectrum with an index of
$\Gamma=1.3$, intrinsic absorption of $N_{\rm H,int}=1.4\times
10^{22}$ cm$^{-2}$, and Galactic absorption of $N_{\rm H,MW}\approx
1.7\times 10^{20}$ cm$^{-2}$ \citep{kbh+05}.  With this model, the
unabsorbed flux is $(5.8\pm 2.0)\times 10^{-15}$ erg s$^{-1}$
cm$^{-2}$ ($0.3-10$ keV).  For a power law
model\footnotemark\footnotetext{This model is appropriate for forward
shock emission with $\nu_c<\nu_X$ and $p=2.45$; see
\S\ref{sec:engine}.} with $\Gamma=2.2$ the resulting flux is only
$\approx 5\%$ lower.  Finally, a multi-temperature accretion disk
blackbody model ({\tt diskbb} in {\tt xspec}) can also fit the data,
with a resulting temperature at the inner disk radius of $kT\approx 1$
keV; thermal disk models with a temperature appropriate to a $\sim
10^6-10^7$ M$_\odot$ supermassive black hole ($kT\lesssim 60$ eV)
cannot reproduce the \chandra\ data (Figure~\ref{fig:xspec}) and
furthermore require an inconsistent radius (see \S\ref{sec:engine}).

\section{Modeling of the Radio Emission}
\label{sec:model}

We model the radio emission from \sw\ following the approach detailed
in Paper I, which is based on the afterglow formulation of
\citet{mgm12} and \citet{gs02}.  For details of the model we refer the
reader to these papers.  For the purpose of estimating the X-ray
emission from the forward shock we also include in the analysis here
the effects of the synchrotron cooling frequency, given by
\citep{gs02}:
\begin{equation}
\nu_c=2.5\times 10^{14}\, \epsilon_{B,-2}^{-3/2}\, L_{\rm
j,iso,48}^{0.5}\, t_{j,6}^{-1}\, n_{18}^{-2}\, \left(\frac{t}{t_j}
\right)^{0.5}\,\, {\rm Hz},
\end{equation}
where $\epsilon_B$ is the fraction of post-shock energy in the
magnetic fields, $L_{\rm j,iso}$ is the kinetic luminosity of the
outflow, $t_j$ is the timescale over which $L_{\rm j,iso}$ is assumed
to be constant (followed by $L_{\rm j,iso}\propto t^{-5/3}$ at $t\ge
t_j$), $n_{18}$ is the circumnuclear density ($n_{\rm CNM}$) at a
fiducial radius of $r=10^{18}$ cm, and we use the notation $X\equiv
10^{y}\,X_{y}$,.  We further assume\footnotemark\footnotetext{Note
that in Paper I we assumed $\epsilon_B=0.1$ and $p=2.5$, which lead to
an overall difference in scaling compared to the results here that can
be determined from the equations in Paper I.  However, the temporal
and radial evolution of the kinetic energy and radial density profile
presented in Paper I remain unchanged.}  that $\epsilon_B=0.01$ and
$p=2.45$.

\begin{figure}
\epsscale{1}
\plotone{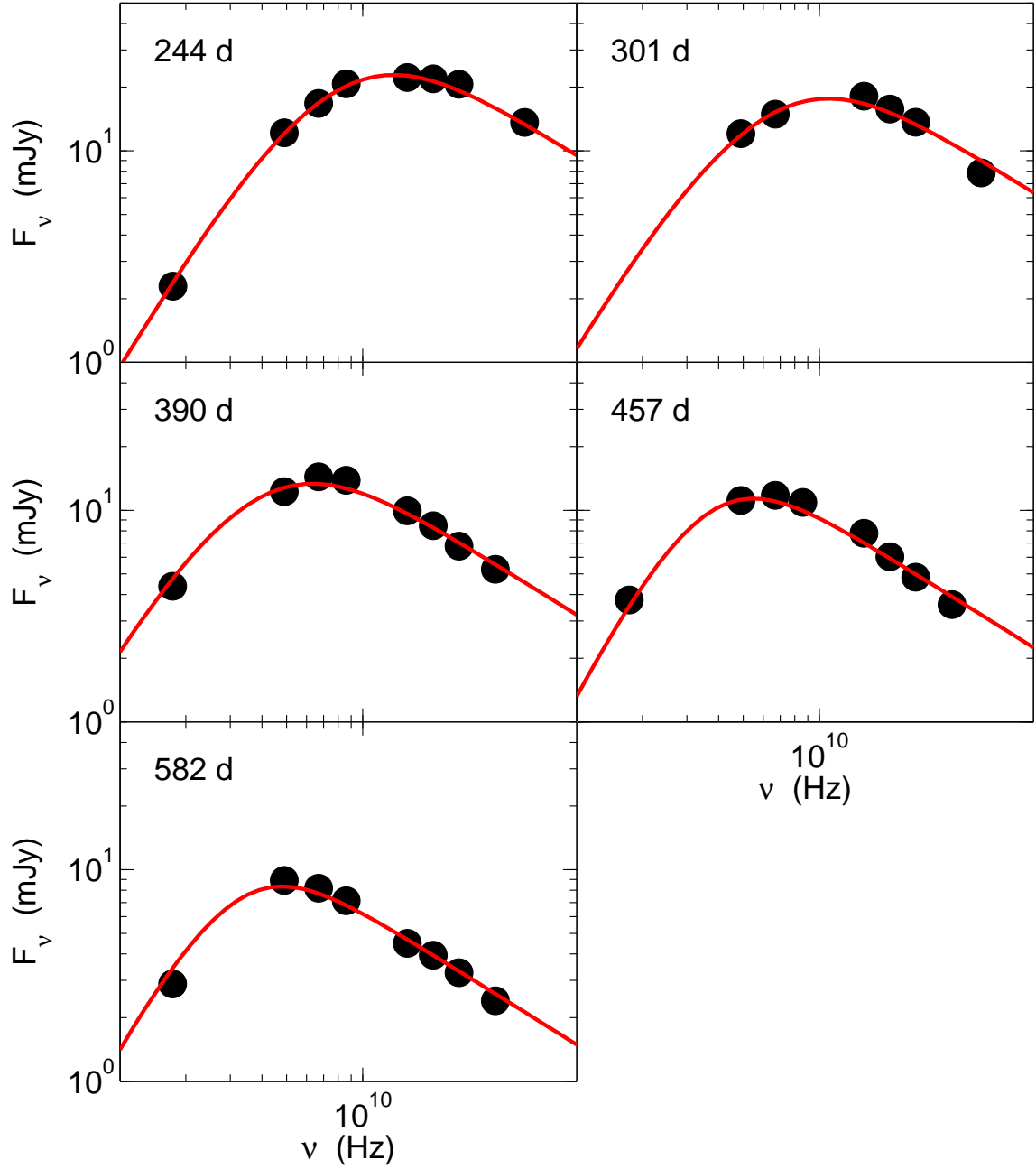}
\caption{Multi-frequency radio spectral energy distributions of \sw\
at $\delta t\approx 244-582$ d.  The solid lines are fits based on the
model described in Paper I, \citet{mgm12}, and \S\ref{sec:model}.  In
each epoch we fit for $L_{\rm j,iso}$ and $n_{18}$ with fixed values
of $\epsilon_e=0.1$, $\epsilon_B=0.01$, and $p=2.45$.
\label{fig:specs}} 
\end{figure}

\begin{figure}
\epsscale{0.65}
\plotone{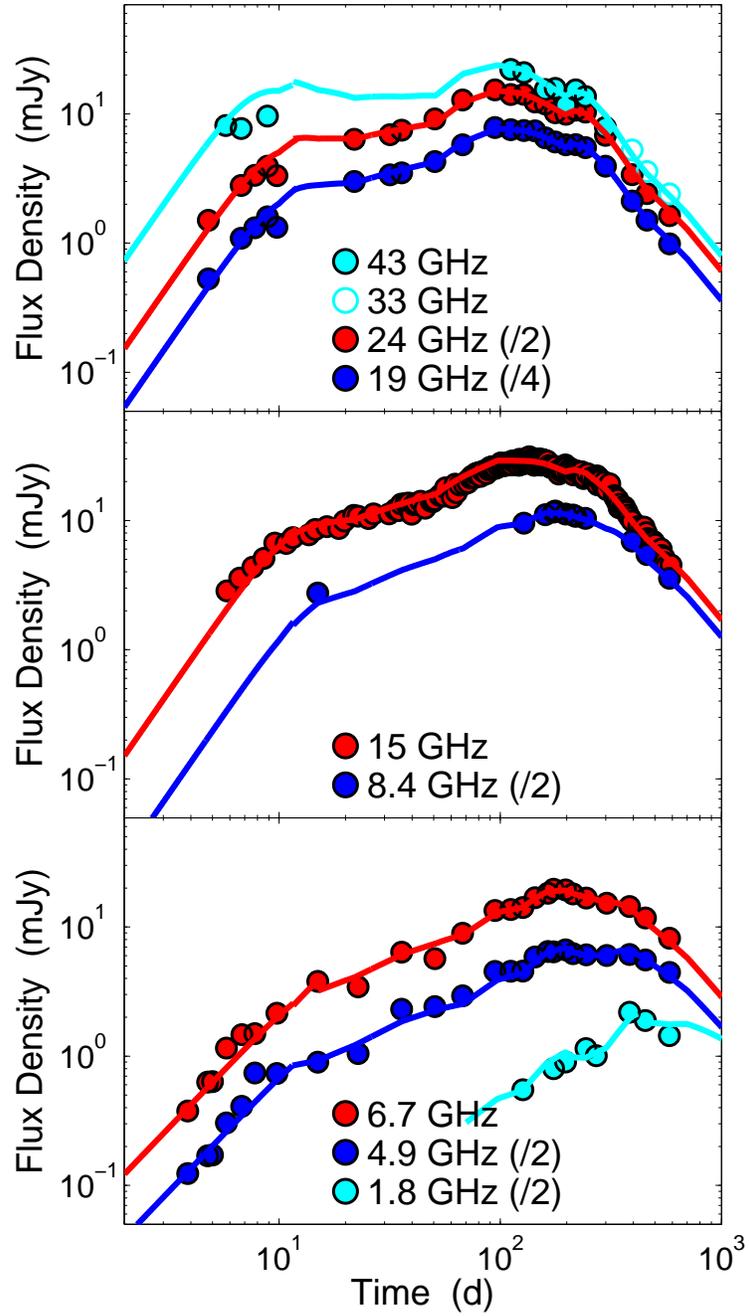}
\caption{Radio light curves of \sw\ extending to $\delta t\approx 600$
d.  The data at $\delta t\approx 5-216$ d were previously presented in
\citet{zbs+11} and Paper I.  The solid lines are models based on
independent fits of broad-band SEDs (Figure~\ref{fig:specs}) using the
model described in Paper I, \citet{mgm12}, and \S\ref{sec:model}.
\label{fig:radio}} 
\end{figure}

As in Paper I, we independently model each broad-band radio spectral
energy distribution (SED) to extract the temporal evolution of the
synchrotron parameters, and in turn the evolution of $L_{\rm j,iso}$,
the emission radius, the jet Lorentz factor ($\Gamma_j$), and the
radial density profile.  The individual SED fits are shown in
Figure~\ref{fig:specs} and the relevant extracted parameters are
listed in Table~\ref{tab:params}.  In Figure~\ref{fig:radio} we plot
the light curves at frequencies of $1.8-43$ GHz, extending to $\approx
600$ d.  Finally, in Figure~\ref{fig:xray} we plot the X-ray data from
\swift/XRT\footnotemark\footnotetext{{\tt
http://www.swift.ac.uk/xrt\_curves/00450158}} and \chandra\ along with
the predicted forward shock emission in the X-ray band based on the
radio SED modeling.

\section{The Relativistic Jet Shuts Off}
\label{sec:engine}

\begin{figure}
\epsscale{1}
\centering
\plotone{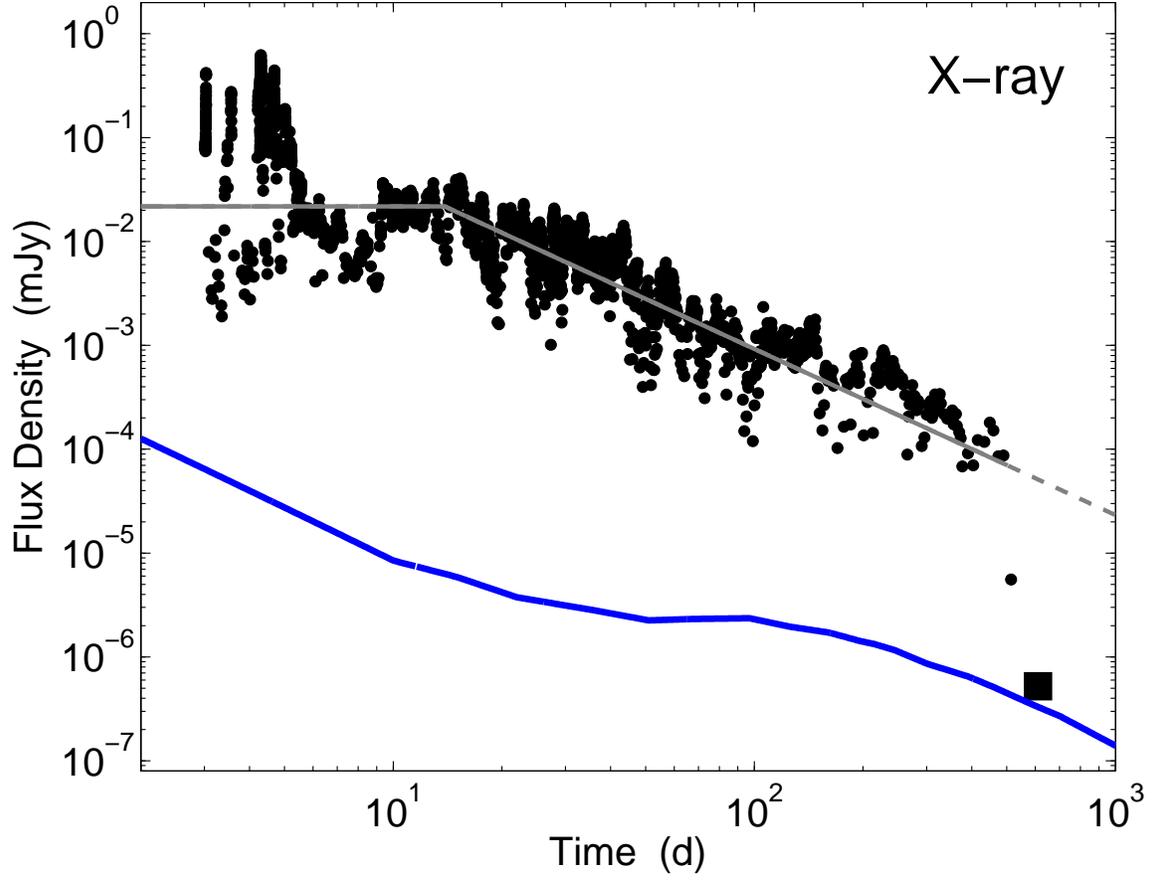}
\caption{X-ray light curve from \swift/XRT (circles) and a late-time
\chandra\ observation (square).  The gray line is a simple model with
a constant flux at $\delta t<t_j$ and $F_X\propto t^{-5/3}$ at $\delta
t\ge t_j$, with $t_j\approx 15$ d.  A rapid decline in the X-ray flux
is evident at $\delta t\gtrsim 500$ d.  The blue line shows the X-ray
emission expected from the forward shock using the synchrotron model
described in \S\ref{sec:model}.  The model indicates that the flux
measured in the \chandra\ observation is consistent with arising from
the forward shock.
\label{fig:xray}} 
\end{figure}

The X-ray light curve at $\delta t\approx 15-500$ d follows a power
law decline, with the expected $F_X\propto t^{-5/3}$
(Figure~\ref{fig:xray}; \citealt{bkg+11}; Paper I).  However, beyond
this point the X-ray flux rapidly declines by a factor of about 15 in
the span of only 25 days, followed by an additional decline of about a
factor of 11 in the subsequent 95 days (see also
\citealt{atel4398,atel4610}).  While the X-ray light curve exhibits
order of magnitude variability in the first few days, followed by
milder variability at later time, a decline by a factor of $\approx
170$ in a narrow span of $\delta t/t\lesssim 0.2$ (and by a factor of
15 in $\delta t/t\approx 0.05$) is unprecedented and points to a
fundamental change in the nature of the emission.  In particular, we
conclude that the mechanism powering the X-ray emission at $\delta
t\lesssim 500$ d has ceased to operate.  The absence of a similar
rapid decline in the radio band supports earlier conclusions that the
radio and X-ray emission arise from distinct physical components
\citep{bgm+11,zbs+11,mgm12}.

In addition, the rapid decline rules out models in which the X-ray
emission at $\delta t\lesssim 500$ d is due to the forward shock or to
reprocessing of radiation by the forward shock, since processes at the
forward shock are expected to occur on a timescale comparable to the
duration of the event, $\delta t/t\approx 1$.  Thus, given the rapid
decline we conclude that the early X-ray emission originated at a
smaller radius than the forward shock, presumably from internal
dissipation in the inner part of the relativistic outflow (at $r\sim
{\rm few}\times 10^{15}$ cm; e.g., \citealt{cgn+12}).  On the other
hand, the low X-ray flux following the steep decline, as measured in
the \chandra\ observation, is fully consistent with emission from the
forward shock at $r\approx 8\times 10^{18}$ cm (Figure~\ref{fig:xray}
and Table~\ref{tab:params}), the same component powering the long-term
radio emission.

An alternative explanation for the low X-ray flux at $\delta t\approx
610$ d is thermal emission from the accretion disk itself.  In this
scenario, for a $3\times 10^6$ M$_\odot$ black hole the effective
temperature is $kT\approx 25$ eV for an inner radius at the tidal
disruption radius, $R_t\approx 12\,R_s\approx 1.1\times 10^{13}$ cm
(e.g., \citealt{ulm99}); here $R_s$ is the Schwarzschild radius.  The
resulting spectral energy distribution severely under-predicts the
observed X-ray flux density, and moreover cannot accommodate the X-ray
spectrum at $\gtrsim 1$ keV due to the expected steep Wien spectrum.
Even a model with a temperature of $kT\approx 60$ eV (corresponding to
an inner disk radius of only $2\,R_s$) cannot accommodate the detected
X-ray emission at $\gtrsim 1$ keV (Figure~\ref{fig:xspec}).  In
particular, for this model to even fit the flux normalization of the
\chandra\ data at $\lesssim 1$ keV requires an inconsistent inner disk
radius of about $40\,R_s$.  A thermal model only fits the data for a
high temperature of $kT\approx 1$ keV, but this is not expected for a
supermassive black hole.  We therefore conclude that forward shock
emission is the most natural explanation for the late-time X-ray flux.

While the nature of relativistic jet generation in TDEs is not fully
understood, an analogy with X-ray binaries suggests that a powerful
jet can be supported as long as the disk is geometrically thick, with
an accretion rate of $\dot{M}\gtrsim \dot{M}_{\rm Edd}$.
\citet{cgn+12} recently presented simulations of the tidal disruption
of a 1 M$_\odot$ star and showed that for a $3\times 10^6$ M$_\odot$
black hole, the peak mass accretion rate is about $240\,\dot{M}_{\rm
Edd}$ (for order unity efficiency), while $\dot{M}\approx \dot{M}_{\rm
Edd}$ at $\delta t\approx 1.5$ yr.  This timescale is remarkably
similar to the time of rapid X-ray decline for \sw, about 370 d in the
rest-frame.  Associating this timescale with an accretion rate of
about $\dot{M}_{\rm Edd}$, we find that the beaming-corrected
\chandra\ X-ray luminosity of $L_X\approx 2\times 10^{42}$ erg
s$^{-1}$ (for $\theta_j=0.1$) is about $0.01\,L_{\rm Edd}$ for a
$3\times 10^6$ M$_\odot$ black hole.  However, the resulting low
efficiency is not surprising given the hard power index of
$\Gamma\approx 1.3$ at $\lesssim 500$ d, which suggests that the bulk
of the energy is radiated above the XRT band.

With the inference that $\dot{M}(\delta t=500\,{\rm d})\approx
\dot{M}_{\rm Edd}\approx 0.006$ M$_\odot$ yr$^{-1}$ we can also
determine the total accreted mass.  Using a simple model with
$\dot{M}(t)=\dot{M}_p$ at $\delta t\lesssim 15$ d and
$\dot{M}(t)=\dot{M}_p\, (t/t_j)^{-5/3}$ at $\delta t\gtrsim 15$ d,
motivated by the X-ray light curve \citep{bkg+11,cgn+12,mgm12}, we
find $\dot{M}_p\approx 350\,\dot{M}_{\rm Edd}$, in good agreement with
the predictions of \citet{cgn+12} for a $3\times 10^6$ M$_\odot$ black
hole.  Integrating the mass accretion rate to $\delta t\approx 370$ d
in the rest-frame, we find a total accreted mass of $\approx 0.15$
M$_\odot$.  This result is consistent with the disruption of a
$\lesssim 1$ M$_\odot$ star.

\begin{figure}
\epsscale{1}
\plotone{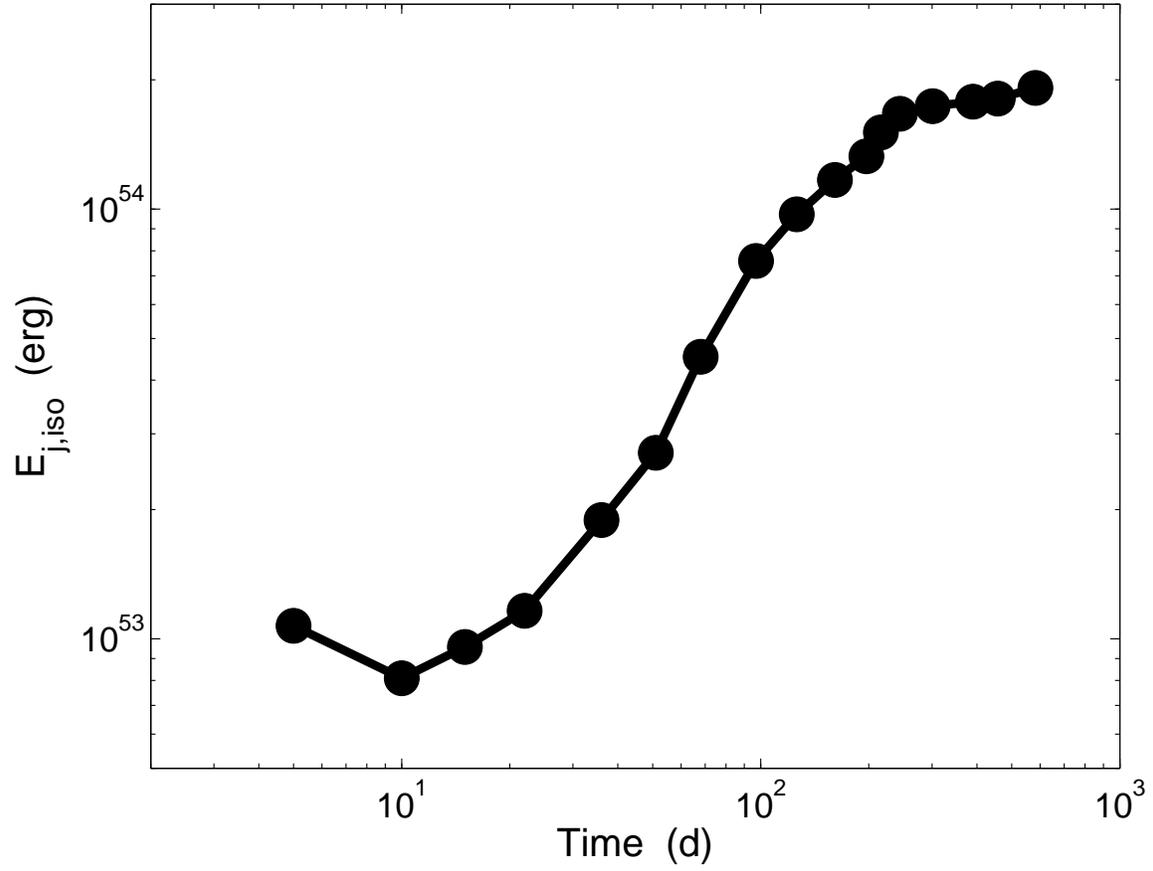}
\caption{Temporal evolution of the isotropic-equivalent integrated
kinetic energy ($E_{\rm j,iso}\equiv L_{\rm j,iso}\,t_j$) based on
modeling of the radio emission (Figure~\ref{fig:specs}).  The rapid
rise at $\delta t\approx 30-250$ d is followed by a mild rise or
plateau to a value of $E_{\rm j,iso}\approx 2\times 10^{54}$ erg.
\label{fig:energy}} 
\end{figure}

In addition to the rapid decline in X-ray emission, which marks the
jet turning off, we also find a change in behavior in the integrated
energy of the forward shock.  Following an increase in $E_{\rm j,iso}$
by about a factor of 15 at $\delta t\approx 30-250$ d, our
measurements at $\delta t\approx 250-600$ d point to a mild rise or a
plateau at a level of $E_{\rm j,iso} \approx 2\times 10^{54}$ erg
(Figure~\ref{fig:energy}).  For an assumed jet opening angle of
$\theta_j\sim 0.1$, this corresponds to a beaming-corrected kinetic
energy of $E_K\approx 10^{52}$ erg.  The flattening in the temporal
evolution of $E_{\rm j,iso}$ is unlikely to be related to the
cessation of jet activity since it begins at an earlier phase.
Instead, it is more likely related to the velocity profile of the
ejecta, as discussed in Paper I, or to a delayed response of the
forward shock to the drop in mass accretion rate below the peak rate
\citep{cgn+12}.  As a result, we expect that the turn off of the
relativistic jet will have only a mild impact on the forward shock
energy, on a timescale of $\delta t\approx 10^3$ d.

\section{Conclusions}
\label{sec:conc}

We present a joint analysis of radio and X-ray observations of \sw\
extending to $\delta t\approx 600$ d.  From the multi-frequency radio
data we determine the integrated energy of the forward shock as a
function of time and find that following an increase in $E_{\rm
j,iso}$ by about a factor of 15 at $\delta t\approx 30-250$ days,
measurements to $\delta t\approx 600$ d reveal a mild rise or plateau
with $E_{\rm j,iso}\approx 2\times 10^{54}$ erg.  X-ray observations
with \swift/XRT and \chandra\ reveal a dramatic change in the light
curve evolution, with a sharp decline by about a factor of 170 at
$\delta t\gtrsim 500-610$ d following a steady $t^{-5/3}$ decline at
$\delta t\approx 15-500$ d.  Using the radio data, we conclude that
the low X-ray flux measured by \chandra\ is consistent with emission
from the forward shock.  The alternative explanation of thermal disk
emission is ruled out by the X-ray flux and spectrum, which instead
require a temperature of $kT\approx 1$ keV, compared to an expected
value of $\lesssim 60$ eV for a supermassive black hole accretion
disk.

The rapid decline suggests that the relativistic jet has turned off,
most likely as a result of a decline in the mass accretion rate below
$\sim \dot{M}_{\rm Edd}$.  With this interpretation, the overall
accreted mass by $\delta t\approx 500$ d is $\approx 0.15$ M$_\odot$
yr$^{-1}$, consistent with the disruption of a solar mass star.
Moreover, the rapid decline, with $\delta t/t\lesssim 0.2$, indicates
that the X-ray emission at $\delta t\lesssim 500$ d did not originate
from the forward shock or from radiation reprocessed by the forward
shock.  Instead it was likely due to internal dissipation in the inner
part of the jet.

Projecting forward, we expect that the X-ray flux evolution will track
the decline rate in the optically-thin high-frequency radio bands with
a potential dispersion of about $\pm 0.25$ due to the response of the
synchrotron cooling frequency to variations in the radial density
profile.  Additional \chandra\ observations in the coming year will
test this prediction.

\acknowledgments We thank Ramesh Narayan and Ryan Chornock for
detailed and helpful discussions.  E.~B.~acknowledges support from the
National Science Foundation through Grant AST-1107973.
A.~M.~S.~acknowledges support from the David and Lucile Packard
Foundation Fellowship for Science and Engineering.  A.~B.~was
supported by a Marie Curie Outgoing International Fellowship (FP7) of
the European Union (project number 275596).  The AMI arrays are
supported by the University of Cambridge and the STFC.  This work made
use of data supplied by the UK Swift Science Data Centre at the
University of Leicester.

$\bibliographystyle{apj}
$\bibliography{journals_apj,refs}

\clearpage
\LongTables
\begin{deluxetable}{rccr}
\tabletypesize{\footnotesize}
\tablecolumns{4} 
\tabcolsep0.15in\footnotesize
\tablewidth{0pt} 
\tablecaption{Radio Observations of \sw\
\label{tab:data}}
\tablehead{
\colhead{$\delta t\,^a$} &
\colhead{Facility}       &
\colhead{Frequency}      &
\colhead{Flux Density}   \\
\colhead{(d)}            &              
\colhead{}               &            
\colhead{(GHz)}          &            
\colhead{(mJy)}            
}
\startdata
  244.23 & JVLA & 1.8 & $2.29\pm 0.08$ \\
  271.95 & JVLA & 1.8 & $2.02\pm 0.23$ \\
  383.92 & JVLA & 1.8 & $4.37\pm 0.10$ \\
  452.66 & JVLA & 1.8 & $3.77\pm 0.09$ \\
  581.31 & JVLA & 1.8 & $2.88\pm 0.08$ \\\hline
  245.23 & JVLA & 4.9 & $12.17\pm 0.05$ \\
  302.95 & JVLA & 4.9 & $12.05\pm 0.05$ \\
  383.92 & JVLA & 4.9 & $12.24\pm 0.03$ \\
  453.66 & JVLA & 4.9 & $11.12\pm 0.03$ \\
  582.31 & JVLA & 4.9 & $ 8.90\pm 0.03$ \\\hline
  245.23 & JVLA & 6.7 & $16.75\pm 0.06$ \\
  302.95 & JVLA & 6.7 & $15.30\pm 0.08$ \\
  383.92 & JVLA & 6.7 & $14.40\pm 0.03$ \\
  453.66 & JVLA & 6.7 & $11.76\pm 0.02$ \\
  582.31 & JVLA & 6.7 & $ 8.18\pm 0.02$ \\\hline
  243.09 & JVLA & 8.6 & $20.76\pm 0.24$ \\
  394.72 & JVLA & 8.6 & $13.84\pm 0.03$ \\
  460.67 & JVLA & 8.6 & $10.89\pm 0.03$ \\
  582.21 & JVLA & 8.6 & $ 7.14\pm 0.03$ \\\hline
  240.25 & AMI-LA & 15.4 & $22.06\pm 0.52$ \\
  247.97 & AMI-LA & 15.4 & $22.99\pm 1.20$ \\
  258.65 & AMI-LA & 15.4 & $21.70\pm 1.03$ \\
  267.86 & AMI-LA & 15.4 & $20.45\pm 0.77$ \\
  270.03 & AMI-LA & 15.4 & $21.60\pm 0.13$ \\
  273.90 & AMI-LA & 15.4 & $22.11\pm 0.38$ \\
  275.76 & AMI-LA & 15.4 & $18.84\pm 0.59$ \\
  278.91 & AMI-LA & 15.4 & $21.38\pm 0.22$ \\
  279.86 & AMI-LA & 15.4 & $20.62\pm 0.36$ \\
  282.87 & AMI-LA & 15.4 & $20.36\pm 0.08$ \\
  289.92 & AMI-LA & 15.4 & $18.68\pm 0.22$ \\
  295.62 & AMI-LA & 15.4 & $20.04\pm 0.45$ \\
  302.98 & AMI-LA & 15.4 & $17.31\pm 0.21$ \\
  312.79 & AMI-LA & 15.4 & $19.21\pm 0.35$ \\
  327.81 & AMI-LA & 15.4 & $15.58\pm 0.14$ \\
  330.81 & AMI-LA & 15.4 & $14.99\pm 0.12$ \\
  332.80 & AMI-LA & 15.4 & $14.94\pm 0.22$ \\
  336.81 & AMI-LA & 15.4 & $14.10\pm 0.22$ \\
  339.73 & AMI-LA & 15.4 & $14.55\pm 0.07$ \\
  344.72 & AMI-LA & 15.4 & $12.53\pm 1.85$ \\
  347.79 & AMI-LA & 15.4 & $13.39\pm 0.07$ \\
  357.53 & AMI-LA & 15.4 & $12.75\pm 1.01$ \\
  364.66 & AMI-LA & 15.4 & $12.66\pm 0.24$ \\
  367.72 & AMI-LA & 15.4 & $12.06\pm 0.05$ \\
  371.68 & AMI-LA & 15.4 & $11.57\pm 0.91$ \\
  373.63 & AMI-LA & 15.4 & $11.76\pm 0.11$ \\
  378.65 & AMI-LA & 15.4 & $10.66\pm 0.69$ \\
  386.44 & AMI-LA & 15.4 & $10.62\pm 0.54$ \\
  394.66 & AMI-LA & 15.4 & $ 9.90\pm 0.11$ \\
  422.84 & AMI-LA & 15.4 & $ 8.92\pm 0.48$ \\
  438.64 & AMI-LA & 15.4 & $ 8.85\pm 0.30$ \\
  444.53 & AMI-LA & 15.4 & $ 8.00\pm 0.36$ \\
  447.52 & AMI-LA & 15.4 & $ 8.87\pm 0.93$ \\
  450.46 & AMI-LA & 15.4 & $ 7.57\pm 0.20$ \\
  457.60 & AMI-LA & 15.4 & $ 8.05\pm 0.20$ \\
  463.22 & AMI-LA & 15.4 & $ 7.09\pm 0.64$ \\
  470.48 & AMI-LA & 15.4 & $ 6.90\pm 0.01$ \\
  477.54 & AMI-LA & 15.4 & $ 6.91\pm 0.55$ \\
  479.47 & AMI-LA & 15.4 & $ 6.36\pm 0.67$ \\
  488.49 & AMI-LA & 15.4 & $ 7.01\pm 0.26$ \\
  492.42 & AMI-LA & 15.4 & $ 6.88\pm 0.37$ \\
  499.30 & AMI-LA & 15.4 & $ 5.78\pm 0.03$ \\
  502.36 & AMI-LA & 15.4 & $ 6.54\pm 0.76$ \\
  513.76 & AMI-LA & 15.4 & $ 6.00\pm 0.77$ \\
  522.32 & AMI-LA & 15.4 & $ 5.38\pm 0.64$ \\
  525.29 & AMI-LA & 15.4 & $ 5.47\pm 0.42$ \\
  528.25 & AMI-LA & 15.4 & $ 5.55\pm 0.20$ \\
  534.40 & AMI-LA & 15.4 & $ 6.00\pm 0.10$ \\
  538.26 & AMI-LA & 15.4 & $ 5.36\pm 0.43$ \\
  550.31 & AMI-LA & 15.4 & $ 5.16\pm 0.35$ \\
  552.13 & AMI-LA & 15.4 & $ 4.72\pm 0.25$ \\
  567.36 & AMI-LA & 15.4 & $ 4.46\pm 0.24$ \\
  592.10 & AMI-LA & 15.4 & $ 4.51\pm 0.27$ \\\hline
  243.09 & JVLA & 19.1 & $21.89\pm 0.10$ \\
  298.96 & JVLA & 19.1 & $15.75\pm 0.06$ \\
  394.72 & JVLA & 19.1 & $ 8.46\pm 0.03$ \\  
  460.67 & JVLA & 19.1 & $ 6.03\pm 0.03$ \\
  582.21 & JVLA & 19.1 & $ 3.94\pm 0.03$ \\\hline
  243.09 & JVLA & 24.4 & $20.65\pm 0.11$ \\
  298.96 & JVLA & 24.4 & $13.64\pm 0.06$ \\
  394.72 & JVLA & 19.1 & $ 6.77\pm 0.04$ \\  
  460.67 & JVLA & 24.4 & $ 4.83\pm 0.03$ \\
  582.21 & JVLA & 24.4 & $ 3.26\pm 0.03$ \\\hline
  394.72 & JVLA & 33.4 & $ 5.26\pm 0.04$ \\
  460.67 & JVLA & 33.4 & $ 3.58\pm 0.04$ \\
  582.21 & JVLA & 33.4 & $ 2.41\pm 0.05$ \\\hline
  243.09 & JVLA & 43.6 & $13.63\pm 0.19$ \\
  298.96 & JVLA & 43.6 & $ 7.86\pm 0.14$ 
\enddata
\tablecomments{$^a$ All values of $\delta t$ are relative to the
initial $\gamma$-ray detection: 2011 March 25.5 UT.}

\end{deluxetable}

\clearpage
\begin{deluxetable}{ccccccccccc}
\tabletypesize{\footnotesize}
\tablecolumns{11} 
\tabcolsep0.1in\footnotesize
\tablewidth{0pt} 
\tablecaption{Results of Broad-band Spectral Energy Distribution Fits
\label{tab:params}}
\tablehead{
\colhead{$\delta t$}                     &
\colhead{${\rm log}(\nu_a)$}             &
\colhead{${\rm log}(\nu_m)$}             &
\colhead{${\rm log}(\nu_c)$}             &
\colhead{${\rm log}(F_{\nu_a})$}         &       
\colhead{${\rm log}(r_{18})$}            &      
\colhead{${\rm log}(\Gamma_{\rm sh})$}   &      
\colhead{${\rm log}(\Gamma_j)$}          &      
\colhead{${\rm log}(L_{\rm j,iso,48})$}  &      
\colhead{${\rm log}(n_{18})$}            &     
\colhead{${\rm log}(n_{\rm CNM})$}       \\      
\colhead{(d)}                            &              
\colhead{(Hz)}                           &            
\colhead{(Hz)}                           &            
\colhead{(Hz)}                           &            
\colhead{(mJy)}                          &           
\colhead{(cm)}                           &                    
\colhead{}                               &   
\colhead{}                               &   
\colhead{(erg s$^{-1}$)}                 &         
\colhead{(cm$^{-3}$)}                    &
\colhead{(cm$^{-3}$)}                    
}
\startdata
244 & $10.04$ & $9.67$ & $13.00$ & $1.99$ & $0.59$ & $0.31$ & $0.35$ & $0.22$ & $1.08$ & $-0.10$ \\
301 & $ 9.96$ & $9.54$ & $13.09$ & $1.92$ & $0.66$ & $0.30$ & $0.33$ & $0.24$ & $1.07$ & $-0.24$ \\
390 & $ 9.71$ & $9.38$ & $13.45$ & $1.72$ & $0.79$ & $0.31$ & $0.33$ & $0.25$ & $0.92$ & $-0.66$ \\
457 & $ 9.62$ & $9.28$ & $13.56$ & $1.64$ & $0.85$ & $0.30$ & $0.33$ & $0.26$ & $0.88$ & $-0.81$ \\
582 & $ 9.58$ & $9.13$ & $13.58$ & $1.60$ & $0.90$ & $0.28$ & $0.30$ & $0.28$ & $0.90$ & $-0.90$
\enddata
\tablecomments{Measured and inferred parameters of the relativistic
outflow and environment of \sw\ from model fits of the individual
multi-frequency SEDs shown in Figure~\ref{fig:specs}.  The model is
described in Paper I, \citet{mgm12}, and \S\ref{sec:model}.}
\end{deluxetable}

\clearpage

\clearpage

\clearpage

\clearpage

\clearpage

\end{document}